\documentclass[12pt]{iopart}

\usepackage{graphicx}
\usepackage{dcolumn}
\usepackage{bm}
\usepackage{graphics,color,epsfig}
\usepackage{subfigure}

\begin{document}

\title{Electron transport through Al-ZnO-Al: an {\it ab initio} calculation}

\author{Zijiang Yang$^1$, Langhui Wan$^1$, Yunjin Yu$^1$, Yadong Wei$^{1,2}$ and Jian Wang$^2$}

\address{$^1$ School of Physics and Institute of
computational condensed matter physics, Shenzhen University,
Shenzhen 518060, P. R. China}
\address{$^2$ Department of Physics, The University of Hong Kong, Pokfulam Road, Hong Kong, P. R. China}

\ead{ywei@szu.edu.cn}

\begin{abstract}
The electron transport properties of ZnO nano-wires coupled by two
aluminium electrodes were studied by {\it ab initio} method based on
non-equilibrium Green's function approach and density functional
theory. A clearly rectifying current-voltage characteristics was
observed. It was found that the contact interfaces between Al-O and
Al-Zn play important roles in the charge transport at low bias
voltage and give very asymmetric I-V characteristics. When the bias
voltage increases, the negative differential resistance occurs at
negative bias voltage. The charge accumulation was calculated and
its behavior was found to be well correlated with the I-V
characteristics. We have also calculated the electrochemical
capacitance which exhibits three plateaus at different bias voltages
which may have potential device application.
\end{abstract}

\pacs{73.40.Sx,72.10.Bg,72.80.Ey}

\submitto{\NT}

\maketitle

\section{Introduction}

The investigations on ZnO material have been continued for many
decades due to its wide range of applications \cite{Ozgur}. Due to
its large direct band gap (${\it E_g}\sim$ 3.3 eV at 300 K) and
large exciton binding energy $\sim$ 60 meV (room-temperature 300 K
is about $\sim$ 25 meV), it is an important material for
manufacturing ultraviolet devices \cite{TangZK} and high temperature
luminescent devices \cite{XuCX, Fischer}. Its easy growth with high
purity makes it a low cost material as high quality substrate
\cite{Look,Ohshima} and its easily etched property makes it possible
to be made as small-size devices. It can be used as space devices
\cite{Kucheyev} and transparent thin-film transistors
\cite{Bae,Kwon} because of its insensitivity to high energy
radiation and visible light. It can also be a candidate as a
spintronic material because Mn-doped ZnO, as well as Fe-, Co-, or
Ni-alloyed ZnO, has high Curie temperature ferromagnetism
\cite{Dietl}. There are many other properties and applications for
ZnO bulk material and ZnO related materials that make ZnO a unique
material for emerging nano-technology \cite{Ozgur}.

Recently, several configurations of ZnO nanostructures such as
nanowires \cite{Heo}, nanotubes \cite{Li,Shi,Riaz}, nanobelts
\cite{Zheng,Jack}, nanosheets \cite{Umar,Suliman} and nanorods
\cite{Li,Riaz}, have attracted much attention due to their potential
applications and easy fabrication \cite{ZLwang}. Such ZnO
nanostructures have novel electrical \cite{Fan, Lao, WangXD} and
optoelectrical \cite{NgHT, Soci} properties as well as mechanical
properties \cite{ZLwang}. They can be used to make sensitive
chemical nanosensors \cite{WanQ,FanZY}, solar cell \cite{Matt},
light-emitting devices \cite{Michae,LiuCH}, nanogenerators
(converting nano-scale mechanical energy into electricity) and
nano-piezotronic devices \cite{ZLwang}. First principles
calculations for the geometry and electronic structure of small ZnO
nanowire and nanobelt were also carried out by many authors
\cite{Jaffe, Erhart, JackY}. More recently, the carrier transport of
ZnO coupled by metal layers such as Au electrode and Mg electrode
were studied by Kamiya \etal \cite{Kamiya}. In their calculation, it
was found that the Au-ZnO-Au interface shows Schottky contact
behavior while the Mg-ZnO-Mg interface shows ohmic contact. The
contact behavior has crucial influence on the device property and is
very sensitive to the fabrication processes \cite{Thaler}. Hence it
is very useful to study the contact behaviors between ZnO and other
materials. In addition, it is very useful to study other transport
properties such as charge polarization and electrochemical
capacitance. In this paper, we concentrated on the transport
behaviors and charge distribution and charge accumulation of
Al-ZnO-Al molecular structure by the density functional theory
combined with nonequilibrium Green's function method. Different from
Au-ZnO-Au and Mg-ZnO-Mg nanostructures, we found that for Al-ZnO-Al
molecular junction the I-V curve is highly asymmetric. In addition,
a negative differential resistance appears at negative bias voltage.
Furthermore, we have calculated the electrochemical capacitance
$C_\mu$ of this device and found that $C_\mu$ depends on bias
voltage in a unique way: there exists three regions where $C_\mu$ is
constant. At small bias, our analysis shows that electrochemical
capacitance is dominated by the classical capacitance whereas at
large bias voltage quantum effect becomes important.

\begin{figure}[b]
\includegraphics*[height=3.3cm,width=8cm]{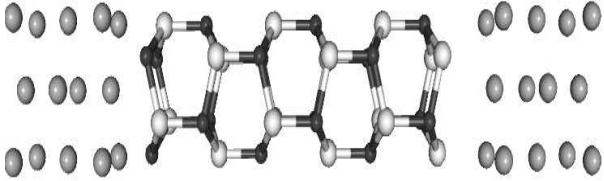}
\caption{Schematic plot of Al-ZnO-Al (Al: gray; Zn: white; O:
black). The direction from left to right is considered as z-axis.}
\label{fig1}
\end{figure}

\section{Method}

\begin{figure}
\includegraphics[height=6cm,width=8cm]{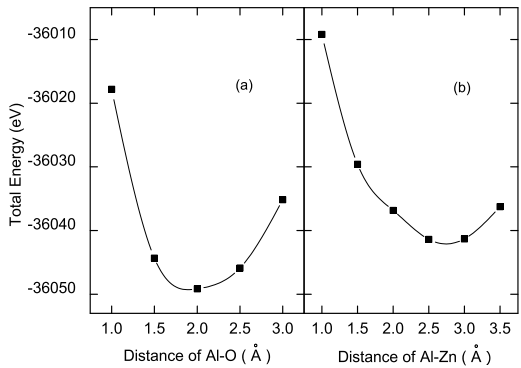}
\caption{Contact distances between aluminium electrode and ZnO
determined by finding the lowest energy. (a) Total energy versus the
distance between Al-O contact; (b) Total energy versus the distance
between Zn-Al contact.}
\label{fig2} 
\end{figure}

Here we present our first principles analysis of electronic
transport through a Al-ZnO-Al nano-device. The calculation package
we used is NADCAL \cite{Guo's Web} (the developed code of McDcal
\cite{Jeremy1}), which is an implementation of real space density
functional theory (DFT) within the Keldysh non-equilibrium Green's
function formalism (NEGF). The DFT method provides the system
electronic structure and NEGF includes information of the
non-equilibrium quantum statistics. The system we studied was a
typical two-probe open system which was divided into a scattering
region and two semi-infinite electrodes. The electronic structure of
the central region was calculated self-consistently and the
couplings between the central region and the two semi-infinite
electrodes were included by the self-energy. Enough buffer layers
for Al atomic electrode were included in the cental region so that
the boundary conditions of the scattering region could be determined
through the corresponding infinite left and right electrodes. We
refer to references \cite{Jeremy1,Jeremy2,Jian1} for detailed
description of NEGF-DFT technique.

Our central scattering region consisted of 6-layers of (0001) ZnO
with width 3.752 {\AA} and three buffer layers of each aluminium
electrodes (see figure \ref{fig1}). Each ZnO layer had 3 oxygen
atoms and 3 zinc atoms. Basis set was composed of s,p,d real space
linear combination atomic orbitals (LCAO). $3d^{10}4s^2$ was chosen
as valence electrons for zinc and $2s^{2}2p^{4}$ for oxygen.
Exchange correlation was treated using LDA\underline{ }PZ
 \cite{LDAPZ}. When solving the Poisson equation in real space one
grid point occupied a volume of $4.583\times10^{-3}$ \AA$^3$. In the
self-consistent calculation, the tolerance was set to be less than
$10^{-4}$ Hartree.

For Al-ZnO-Al structure in figure \ref{fig1}, the left electrode was
coupled directly to oxygen atoms (Al-O contact) and the right
electrode was coupled to zinc atoms (Al-Zn contact). Instead of
relaxing ZnO and aluminium electrodes, we determined the distances
between ZnO and electrodes by finding the minimum of total energy
when changing the distance of Al-O and Zn-Al respectively. According
to figure \ref{fig2} we set the distance between aluminum electrode
and oxygen side of ZnO to be 1.95 {\AA} while the distance of Zn-Al
to be 2.75 \AA.

For the two-probe NEGF+DFT formalism, the electronic current can be
obtained by Landauer-B$\ddot{u}$ttiker formula
\begin{equation}
I=\frac{2q}{h} \int dE
{\Tr}[\hat{T}(E,V_L,V_R)](f_L(E,\mu_L)-f_R(E,\mu_R)),
\label{ivcurve}
\end{equation}
where $\hat{T}(E,V_L,V_R)$ is the transmission matrix and is
obtained by Green's function from the following formula
\begin{equation}
 \hat{T}(E,V_L,V_R)=\Gamma_L G^r \Gamma_R G^a. \label{transT}
\end{equation}
$f_{L/R}=[1+\exp((E-\mu_{L/R})/kT)]^{-1}$ is the Fermi distribution
function of left/right electron reservoir connected to the
left/right electrode at infinite $z=-\infty/\infty$.
$\mu_{L/R}=\mu_{l/r}+qV_{L/R}$ is the electrochemical potential with
$\mu_{l/r}$ the chemical potential of left/right lead and $V_{L/R}$
the applied potential on the left/right lead.

\section{Results and Discussions}

\begin{figure} [b]
\includegraphics[height=8cm, width=8cm]{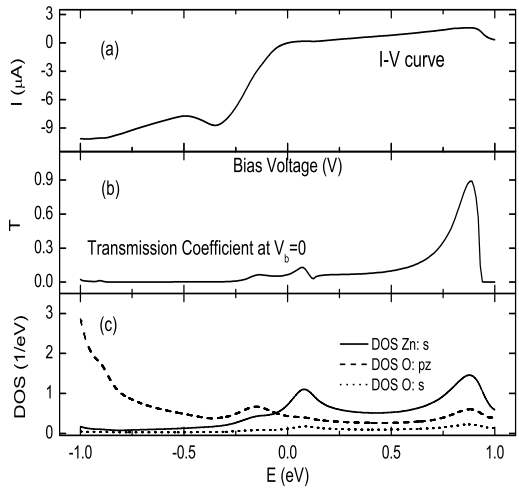}
\caption{(a)Current as a function of bias voltage $V_b=V_L-V_R$;
(b)Transmission coefficient versus energy at bias voltage $V_b=0$;
(c) DOS versus energy due to zinc $4s$ state (Zn:s solid line),
oxygen $2p$ state in $z$-direction (O:pz dashed line) and oxygen
$2s$ state (O:s dotted line)}
\label{fig3} 
\end{figure}

\begin{figure}[b]
\includegraphics[height=6.5cm, width=8cm]{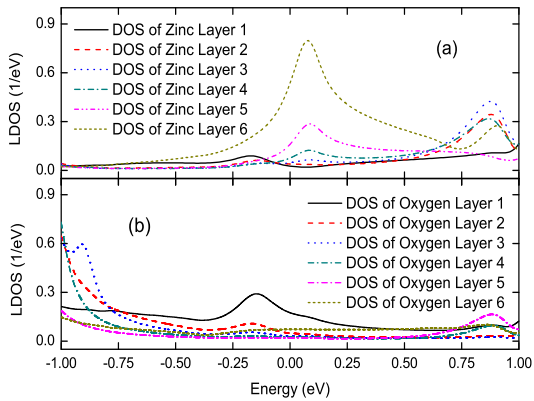}
\caption{(a) Local density of states of zinc $4s$-state on each
layer; (b) Local density of states of oxygen $p_z$-state on each
layer.}
\label{fig4} 
\end{figure}

In figure \ref{fig3}, we plotted the current-voltage (I-V) curve of
Al-ZnO-Al. A rectifying I-V characteristics is clearly observed. The
current increases slowly with increasing of the positive bias
voltage $V_b$ but decreases quickly when the bias voltage reverses
its sign. Different behaviors of I-V characteristics when the bias
voltage changes from positive to negative can be understood as
follows. At zero temperature and in hartree unit, equation
(\ref{ivcurve}) becomes (We set $V_R=0$ in our calculation.)
\begin{equation}
I=2 \int_{-V_b}^{0} \frac{dE}{2\pi}{\Tr}\hat{T}(E,V_b).
\label{ivcurve1}
\end{equation}

We find that when the positive bias voltage is applied, the current
is contributed from the energy integral interval $E=(-|V_b|,0)$. But
for negative voltage $V_b<0$, the transmission coefficient will be
integrated on the positive interval $E=(0,|V_b|)$. Although the
transmission coefficient is a function of the bias voltage, we first
analyze the I-V curve near $V_b=0$. For this purpose, we plotted the
transmission coefficient and the density of states (DOS) versus the
Fermi energy at $V_b=0$ in figure \ref{fig3}. In our calculation we
find that three kinds of states give major contributions to the I-V
curve. They are zinc $4s$-orbital, oxygen $z$-component of
$2p$-orbital and oxygen $2s$-orbital. Among them, contribution due
to oxygen $2s$-orbital is the smallest so we focus on the other two
orbits. It is clear that when $E>0$ the DOS of zinc $s$-orbital is
much larger than that of oxygen $p_z$-orbital and when $E<0$ both
zinc $s$-orbital and oxygen $p_z$-orbital contribute near $E=0$ but
the contribution is less than that of the zinc $s$-state at $E>0$.
So we conclude that when negative bias voltage is added electrons
pass device mainly from zinc $s$-orbital but when the positive bias
voltage is added electrons pass device mainly from both zinc
$s$-orbital and oxygen $p_z$-orbital. To provide further evidence,
we analyze the local density of states (LDOS) according to the
layers of ZnO in the scattering region (see figure \ref{fig4}). Note
that we label the layers of oxygen and zinc from left to right. One
can find that each layer has very different LDOS. For zinc atoms,
the rightmost layer (the layer 6, i.e. the layer coupled directly to
aluminium lead) gives the largest LDOS around the Fermi level. (The
Fermi level is shifted to $E=0$ in our calculation.) When the energy
increases the LDOS of middle layers (layers 2 to 4) gradually
increase and reach their maxima at about $E=0.83$ eV. But the LDOS
of all zinc layers become small when $E<-0.25$ eV. For oxygen atoms,
similar result is found if we recall that the first oxygen layer is
connected directly to the left aluminium electrode. The LDOS of
oxygen give the largest value at the first oxygen layer around the
Fermi level $E=0$. When $E$ becomes quite small and closes to
$E=-1.0$ eV, the LDOS of the middle layers (layer 3 and layer 4)
dominate and when $E>0$ the LDOS of oxygen are small compared with
the those of $E<0$. In another word, when small negative bias
voltage is applied the contact interface between Al-Zn dominates I-V
behavior. But when the positive bias voltage is applied, both Al-Zn
interface and Al-O interface will dominate the charge transport. The
rectifying behavior is mainly controlled by the contact interface
characteristics due to the different coupling atoms Al-Zn and Al-O
when bias voltage is small.

\begin{figure} [t]
  \includegraphics[height=6cm,width=7cm]{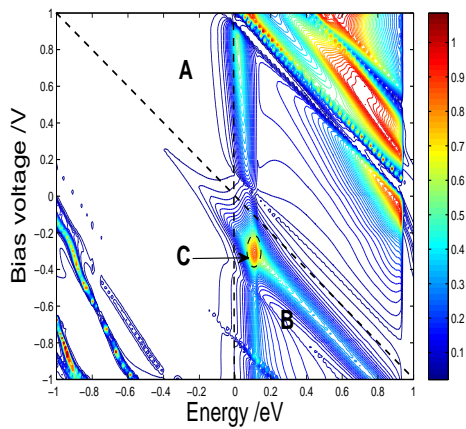}
  \caption{Transmission coefficient versus bias voltage $V_b$ and
  energy $E$. The dashed black lines are only used as symbol lines
  in order to depict the region A, B and C. }
  \label{fig5} 
\end{figure}

Actually the transmission coefficient $T={\Tr}[\hat{T}(E,V_b)]$ is a
function of the bias voltage $V_b=V_L-V_R$ at zero temperature so
the situation becomes more complicated at large $V_b$. In figure
\ref{fig5}, the transmission coefficient as a function of the energy
$E$ and the bias voltage $V_b$ was depicted. The I-V curve for
positive bias voltage corresponds to the triangular area A between
the vertical line $E=0$ and the slope line $V_b=-E$ for bias voltage
$V_b>0$ and energy $E<0$ (see the dashed black lines in figure
\ref{fig5}) and the I-V curve for negative bias voltage corresponds
to the triangular area B between the above two lines for $V_b<0$ and
$E>0$. We find that in the right low triangular area B the
transmission coefficient is much large within the interval
$V_b\in(-0.4,-0.2)$ V and reach maximum around $V_b\sim-0.35$ V (see
the C region in figure \ref{fig5}). That's why the absolute current
value decreases when the absolute value of bias voltage increases
around $V_b\in(-0.5,-0.35)$ V and the negative differential
resistance appears in that voltage region (see I-V curve on the
interval $(-0.5,-0.35)$ V in figure \ref{fig3}(a)).

\begin{figure}
  \includegraphics[height=6.5cm,width=8cm]{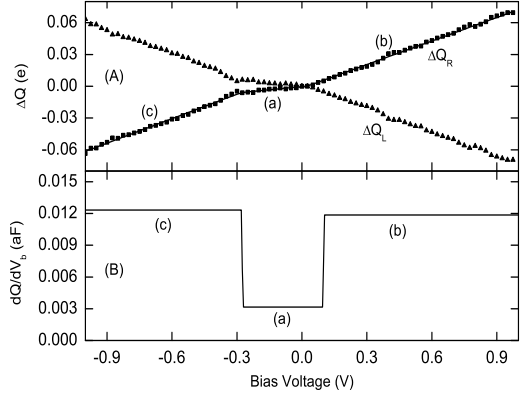}
  \caption{(A)Charge accumulations versus bias voltage
  on left scattering region ($\Delta Q_L$: triangles) and right
  scattering region ($\Delta Q_R$: squares). The black solid
  line  $Q(V_b)$ is the fitted curve of $\Delta Q_R$; (B) Electrochemical
  differential capacitance $dQ/dV_b$ versus bias voltage.}
  \label{fig6} 
\end{figure}

For the quantum devices, due to finite density of states the charge
accumulation in the scattering region differs from the charge
accumulation in the classical metal capacitance and one should use
the concept of electrochemical capacitance $C_{\mu}=\Delta Q /
\Delta \mu$, where $\Delta\mu$ is the electrochemical potential
difference between the two electron reservoirs \cite{Buttiker2,
Jian2,ywei2} and the $\Delta Q$ is the charge accumulation. For our
two-probe problem, both electrodes are aluminium so that
$\mu_l=\mu_r$. Hence, at zero temperature, we have
$\Delta\mu=V_b=V_L-V_R$. When the bias voltage is applied, the
electrochemical potential across two probes is changed by
$\Delta\mu=V_b$ from equilibrium, charges will be injected into the
scattering region. The induced charges will occur due to the
electron-electron interaction. In DFT, the injected and induced
charges are automatically considered by solving the Kohn-Sham
equation and the Poisson equation self-consistently and the Gauge
invariant is also satisfied automatically \cite{Jian3}. Such a
charge accumulation is closely related to the I-V characteristics
and can only be determined by solving scattering problem. For a
rough estimate, we divide the system into two parts from the middle
of scattering region called left region and right region. We then
calculate the charge accumulation $\Delta Q_L(V_b)=Q_L(V_b)-Q_L(0)$
and $\Delta Q_R(V_b)=Q_R(V_b)-Q_R(0)$ and plot them in figure
\ref{fig6}. The electrochemical differential capacitance
$C_{\mu}=dQ/dV_b$ was also plotted in figure \ref{fig6}. Here, $Q$
is the fitted curve of $\Delta Q_R(V_b)$. Following observations are
in order. First of all, the electrochemical capacitance depends on
the voltage although it is a piece-wise constant in three regions
labeled as (a), (b) and (c). Secondly, the asymmetry of
electrochemical capacitance $C_\mu$ is clearly seen. In region (a)
which is asymmetric with respect to the bias voltage
$V_b=(-0.27,0.01)$ V the electrochemical capacitance assumes a
smaller constant value $C_\mu=0.003$ aF. While in region (b) and
(c), $C_\mu$ are approximately four times larger. The value of the
electrochemical capacitance also exhibits asymmetry between region
(b) and (c). Finally, the behavior of electrochemical capacitance in
region (a) can be understood {\it qualitatively} as follow. Since in
region (a) the current is small, we can use the following formula
for electrochemical capacitance (where the current is assumed to be
zero) \cite{but1}
\begin{equation}
\frac{1}{C_\mu} = \frac{1}{C_0}+ \frac{1}{e^2 D_L}+\frac{1}{e^2
D_{R}}, \label{eq1}
\end{equation}
where $C_0$ is the classical capacitance which can be roughly
estimated using the parallel plate capacitor $C_0 = \epsilon_0 A/d
\sim 0.002$ aF where A is the cross-section area and d is the
separation of two plates. Here $D_L$ and $D_{R}$ are the DOS of the
left and right regions, respectively. From figure \ref{fig4}, we see
that $D_L$ and $D_R$ are much larger than $C_0/e^2$ and hence can be
neglected in equation (\ref{eq1}). Therefore, we conclude that in
region (a), the electrochemical capacitance is dominated by the
classical capacitance. In region (b) and (c), however, quantum
effect must be considered. At the moment, there is no analytic
formula to account for the electrochemical capacitance at the
nonlinear bias voltage.

In summary, we have analyzed the electron transport through the
Al-ZnO-Al atomic junction and clear rectifying characteristic was
observed. It was found that the contact interfaces between aluminium
electrodes and central device ZnO play essential roles in the
current-voltage behaviors when the bias voltage is small. When the
negative bias voltage is applied, the major contribution to the
electron transport comes from Zn $s$-states. But when the positive
bias voltage is applied, both Zn $s$-state and O $p_z$-state are
responsible in the charge transport. The negative differential
resistance was found in our device. Our results showed that for such
a quantum open system the charge accumulations in the scattering
region are closely correlated with the applied bias voltage and I-V
characteristics. The electrochemical capacitance was found to have
three plateaus. At small bias, the electrochemical capacitance is
dominated by the classical capacitance. At large bias, however,
quantum effect becomes important. Our findings may have potential
device application for future nano-technology.

\ack{Acknowledgments} We gratefully acknowledge support by a RGC
grant from the Government of HKSAP under grant number HKU 704308P
and the grant from the NSFC under grant number 10947018.


\end{document}